\documentclass[pdflatex,sn-mathphys-num]{sn-jnl}

\usepackage{graphicx}%
\usepackage{multirow}%
\usepackage{amsmath,amssymb,amsfonts}%
\usepackage{amsthm}%
\usepackage{mathrsfs}%
\usepackage[title]{appendix}%
\usepackage{xcolor}%
\usepackage{textcomp}%
\usepackage{manyfoot}%
\usepackage{booktabs}%
\usepackage{algorithm}%
\usepackage{algorithmicx}%
\usepackage{algpseudocode}%
\usepackage{listings}%
\usepackage{siunitx}
\usepackage{threeparttable}
\usepackage{float}
\usepackage{placeins}

\theoremstyle{thmstyleone}%
\theoremstyle{thmstyletwo}%

\theoremstyle{thmstylethree}%

\begin{document}

\title[Article Title]{Policy, Risk, and Norms Shape Collective Behaviors Worldwide}


\author*[1]{\fnm{Dhruv } \sur{Mittal}}\email{d.mittal@uva.nl}

\author[2,3]{\fnm{Sara M. } \sur{Constantino}}

\author[4]{\fnm{Simon A.} \sur{Levin}}

\author[1]{\fnm{Peter} \sur{Sloot}}

\author[5]{\fnm{Elke U.} \sur{Weber}}

\author[1,6]{\fnm{Vítor V. } \sur{Vasconcelos}}

\affil[1]{\orgdiv{Computational Science Lab, Informatics Institute}, \orgname{University of Amsterdam}, \orgaddress{\city{Amsterdam}, \country{Netherlands}}}
\affil[2]{\orgdiv{Doerr School of Sustainability}, \orgname{Stanford University}, \orgaddress{Stanford, CA 94305}}
\affil[3]{\orgdiv{School of Public and International Affairs}, \orgname{Princeton University}, \orgaddress{Princeton, NJ 08544}}
\affil[4]{\orgdiv{Department of Ecology and Evolutionary Biology}, \orgname{Princeton University}, \orgaddress{Princeton, NJ 08544}}
\affil[5]{\orgdiv{Andlinger Center for Energy and the Environment}, \orgname{Princeton University}, \orgaddress{Princeton, NJ 08544}}

\affil[6]{\orgdiv{POLDER, Institute for Advanced Study}, \orgname{University of Amsterdam}, \orgaddress{\city{Amsterdam}, \country{Netherlands}}}


\abstract{
Societal responses to environmental change vary widely, even under comparable shocks, reflecting differences in both policy measures and public reactions shaped by cultural and socioeconomic contexts. We examine mask-wearing dynamics across 47 countries during the COVID-19 pandemic using a process-based, utility-driven model of individual behavior with three evolving drivers: policy stringency, disease risk, and social norms to understand emergent collective behavior. Calibrated with daily data on mask usage, COVID-19 deaths, and policy mandates, the model reproduces diverse national trajectories with minimal complexity. Policy and norms are crucial for explaining variation, and we find significant associations between weights for all three drivers and cultural and socioeconomic indicators. Our findings demonstrate how mechanistic models can uncover the processes shaping collective behavior, enabling policymakers to anticipate the magnitude and timing of behavioral change and design more effective, context-sensitive interventions.


}





\maketitle

\section{Introduction}\label{sec1}


The COVID-19 pandemic offered a powerful, real-time demonstration of how coordinated collective action---in the form of mask-wearing, social distancing, and vaccination uptake---can significantly mitigate loss of life and disease transmission \cite{ aravindakshan2022impact, qiu2022understanding}. This global event also illustrated a more nuanced principle: individual decisions can drive large-scale outcomes, especially when social norms exert a strong influence on behavior \cite{castilla2017social,farahbakhsh2024tipping}.

Yet, the connection between individual decisions and their potentially large societal and environmental impacts is often obscured by spatial and temporal scales, making it challenging for individuals to identify and adopt optimal behaviors \cite{levin2013social}. Furthermore, collective behaviors themselves are shaped by inherent social feedback mechanisms, especially in uncertain or rapidly changing environments. In such contexts, individuals rely heavily on social cues, observing and adopting the beliefs and behaviors exhibited by those around them \cite{heiman2023descriptive, bokemper2021experimental, mladenovic2023influence}. While such social conformity can reinforce existing patterns of behaviors and beliefs, it also holds the potential for rapid and widespread social change (e.g., social tipping), especially when behaviors are highly visible or intentionally emphasized through public policy and media \cite{nyborg2016social,gavrilets2020dynamics, mittal2024anti,syropoulos2024expressive}.

Effective governance of these complex dynamics thus requires strategic intervention, often through targeted information campaigns and policy measures, that are designed with an understanding of these social processes \cite{constantino2021source, baxter2022local,constantino2022scaling,nyborg2016social}. The efficacy of interventions hinges on how populations and social dynamics respond to these institutional signals. Prior research has shown that cultural values and socioeconomic conditions can influence how individuals process information, perceive risks, adhere to social norms, and respond to different government policies \cite{weber2006experience, yang2022sociocultural, wei2023people}. Recognizing and leveraging cultural, social, and psychological influences on individual and collective behavior is essential for policymakers and institutions aiming to induce widespread beneficial behavioral and social change \cite{bavel2020using}.

\begin{figure}[h]
\centering
\includegraphics[width=0.6\textwidth]{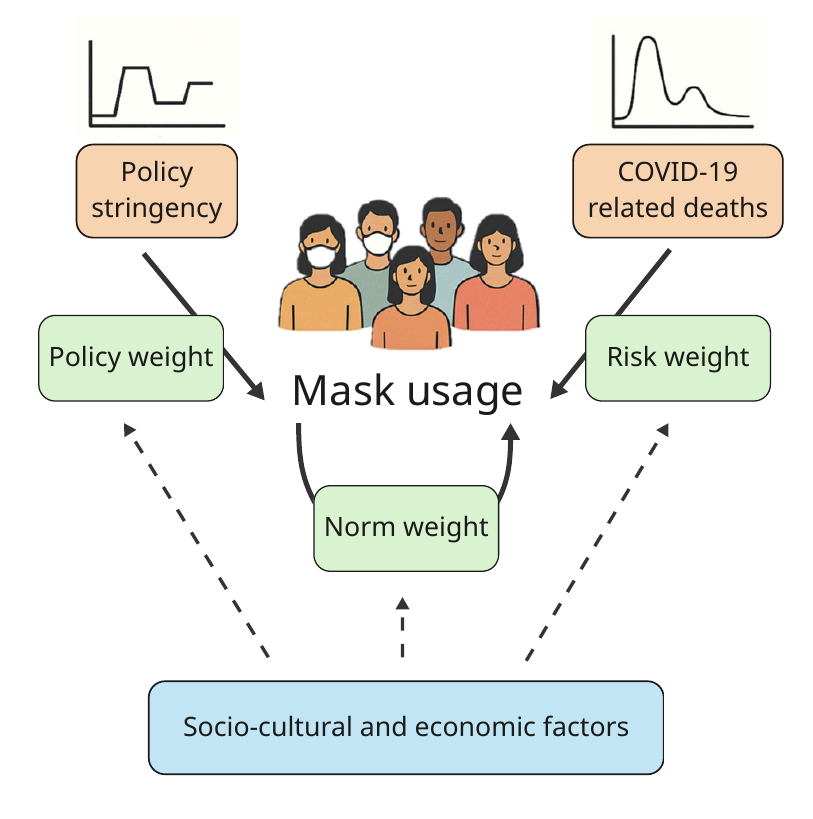}
\caption{\textbf{Conceptual illustration of the model.} Individuals make decisions about wearing masks by considering government policy (i.e., evolving mask mandates), personal health risks (i.e., based on daily COVID-19-related deaths), and observation of their peers (i.e., local social norms). The two data inputs to the model---policy stringency and COVID-19-related fatalities per million---are highlighted in red. We calibrate the weights, modulating the relative effect of the three behavioral mechanisms (highlighted in green), against the UMD-CTIS survey for 47 countries. Details can be found in the Data and Methods section. }\label{fig1}
\end{figure}


Historically, the empirical calibration of theoretical models to real-world behavioral data has been hampered by the lack of high-resolution temporal and spatial datasets linking local behavioral factors and attitudes to collective outcomes \cite{lazer2009computational}. As a result, much of the literature has relied on cross-sectional snapshots and simple regression-based analyses \cite{gelfand2021relationship,lu2021collectivism}, which can fail to capture the dynamic processes underlying behavior or worse produce misleading interpretations \cite{flache2017models}. With the growing availability of longitudinal and fine-grained behavioral data, there is now an opportunity—and a need—to move beyond purely correlational approaches. Mechanistic models that explicitly represent processes such as social feedback and adaptive responses to changing external conditions are essential to leverage these data fully and to deepen our understanding of the drivers of behavioral dynamics \cite{holme2015mechanistic}.

Here, we develop such a model of mask-wearing behavior during the COVID-19 pandemic---a widely recognized, effective non-pharmaceutical intervention for controlling infection spread\cite{aravindakshan2022impact,chen2025adherence}. The adoption of masks varied significantly both across and within countries due to differing policies, emergent social norms, and evolving risk conditions \cite{randler2025covid, baxter2022local}. In some countries, mask uptake was swift, widespread, and persistent, while in others it was slow, inconsistent, or persistently low. Previous research indicates that cultural dimensions, such as collectivism and tightness–looseness, significantly influenced these cross-national differences \cite{lu2021collectivism, gelfand2011differences, yang2022sociocultural}. However, a notable gap remains in our understanding of how exogenous drivers, such as evolving mask-wearing policies and changing disease risks, interact with the endogenous process of local norm formation to shape individual and collective behavioral trajectories at the country level \cite{wismans2022face}.

Prior studies have emphasized descriptive norms, policy stringency, and individual risk perception as crucial determinants of mask-wearing behavior in local contexts \cite{heiman2023descriptive, bokemper2021experimental, wismans2022face}. Yet, a systematic exploration of empirically informed parsimonious models that consider how these factors collectively shape behavioral dynamics across diverse socio-cultural contexts is limited. Addressing this research gap, we analyze publicly available data of mask-wearing behavior, policy interventions, and COVID-19 incidence across 47 countries at daily resolution \cite{hale2021global, wang2024exploring,fan2020university}. Through this analysis, we seek to elucidate how socio-cultural contexts mediate collective behavioral responses to evolving exogenous factors. \cite{badillo2021global}.

\begin{figure}[t]
\centering
\includegraphics[width=1\textwidth]{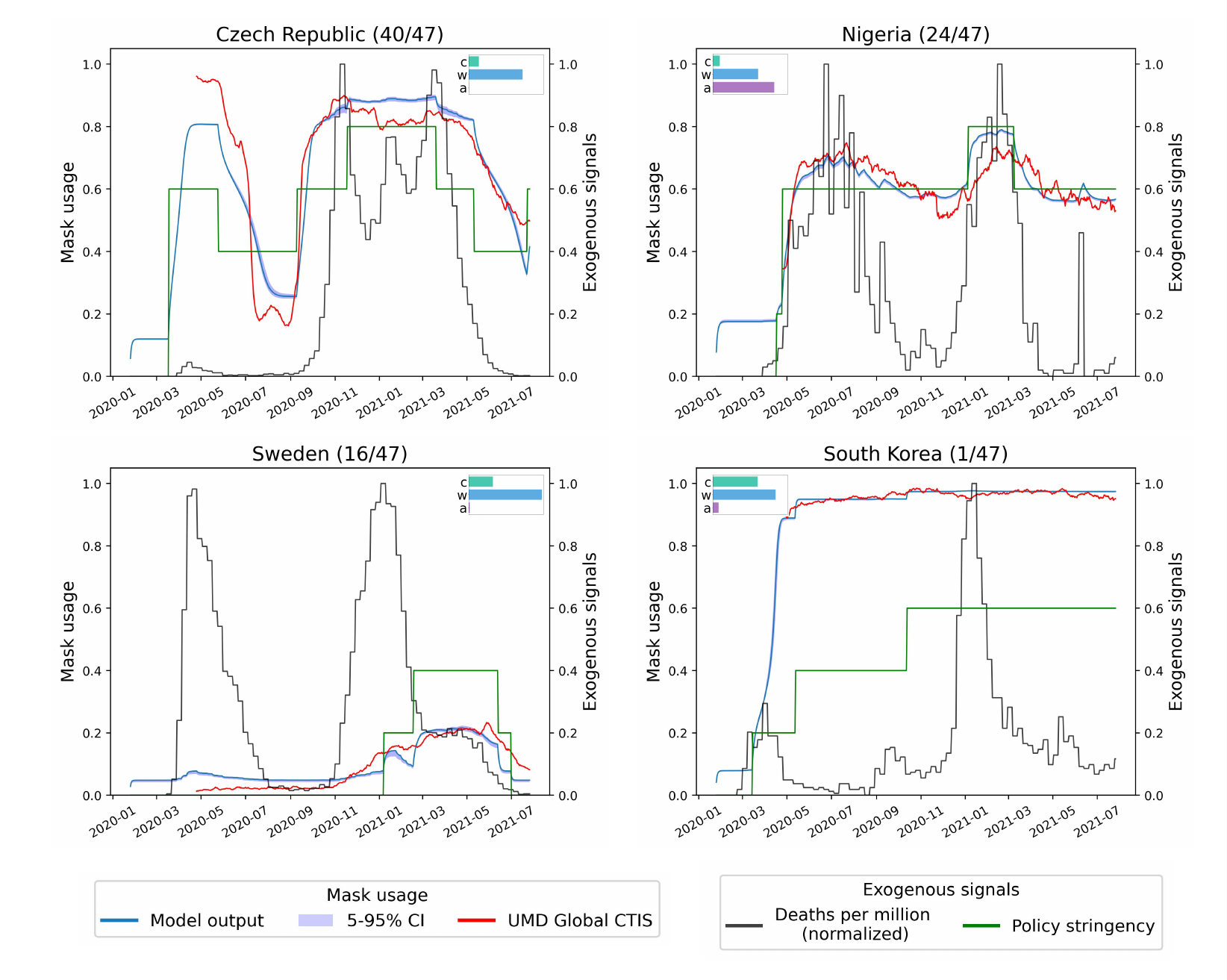}
\caption{\textbf{Recovering mask-wearing dynamics in highly varied case studies:} The mask usage data from UMD Global CTIS and the calibrated model output are plotted along with COVID-19 deaths and the stringency of masking policy for 4 illustrative countries: the Czech Republic, Nigeria, Sweden, and South Korea. High variability in disease progression and policy stringency is illustrated across countries. The confidence intervals of the model output are plotted based on the parameter uncertainty calculated by bootstrapping the residuals. The country's rank in terms of model error is denoted next to the country name (high rank means low model error). Inset bar graphs show normalized values of policy weight ($c$), norm weight ($w$), and risk weight($a$), highlighting the different drivers of mask usage across country contexts. Notably, Nigerian residents appear to place a high weight on risk, while this is not the case in European countries. South Koreans place a greater weight on policy signals. }\label{fig2}
\end{figure}

Objective risks and policies alone cannot explain the different trajectories across countries. The reason is that these signals are mediated via cultural, social, and perceptual factors. Behavior, particularly in uncertain contexts, is socially interdependent, leading small initial changes in behavior (e.g., due to policy measures or local shocks) to spread, emerging as new norms. However, these dynamics, and the weight placed on different inputs to decision-making, depend on the cultural context \cite{mladenovic2023influence, randler2025covid}. Mask-wearing is an illustrative case of a behavior that offers both individual protection and positive externalities or collective benefits. Accordingly, populations in collectivist cultures generally demonstrated greater mask adherence, whereas in other contexts, mask mandates were seen as infringements on personal freedoms \cite {traulsen2023individual,lu2021collectivism, baxter2022local}.

To explore these complex interactions systematically, we employ an individual-level model of decision-making that integrates external evolving institutional and environmental signals with social feedback mechanisms in the form of observed norms. Using a representative agent for each country, we understand collective mask-wearing dynamics in 47 countries \cite{qiu2022understanding}. We expect that the decision weights being applied to risk, policy, and social signals will depend on the cultural and socioeconomic context. We estimate weights given to different signals by calibrating the model with granular data on evolving institutional signals (i.e., national-level mask mandates) and risk conditions (i.e., COVID-19-related deaths per million) (Figure \ref{fig1}). The calibration results indicate that the weights given to policy, risk, and social factors vary systematically across countries \cite{wang2024exploring, heiman2023descriptive}. The results also suggest that while policy was the main driver, social norms and risk are important to accurately capture observed mask-wearing dynamics across countries. Further, analysis reveals significant associations between the estimated weights and broader socio-cultural and economic (SCE) factors indicators indicating how these factors could have modulated the behavioral trajectory.

\section{Results}\label{sec2}

We develop a utility-based framework to model individual mask-wearing decisions. Individuals consider the institutional signal (policy stringency), risk (COVID-19-related deaths), and descriptive norm (observed mask usage) when deciding whether to wear a mask. Conformity to descriptive norms results in benefits from coordinating with others (e.g., the avoidance of social sanctions). In this setup, individuals influence each other in an evolving environment of policy and risk. This results in emergent collective behavioral dynamics captured by the fraction of individuals in country i, wearing a mask at time $t$, \(x_t\). The utility an individual considers is:
\[
U_t = -k + a\, I_t + w_i\,(x_t - 0.5) + c_i\, S_t, \tag{1}
\label{eq1}
\]
where \(I_t\) denotes COVID-19–related deaths per million at time \(t\), and \(S_t\) represents policy stringency at time \(t\)(on a scale of 1-5), both evolving exogenous factors specific to country i. In this formulation, the parameters \(a\), \(c\), and \(w\) capture the weight given to the three signals, relative to \(k\), which is the constant baseline resistance to mask-wearing (fixed at 1 across countries). At each time step (day), a fraction of the population ($\mu$) considers changing behavior based on the calculated utility of wearing a mask, with $\beta$ indicating the level of rationality, capturing the steepness of the response function.

We assume that individuals in a country share the same decision weights, thus assuming a representative agent for the country. Further, we assume that the agents are well-mixed, i.e., that every individual observes the same mask usage. These assumptions allow us to take a mean-field approach and calculate the rate of change in the expected collective behavior (see Eq. \ref{eq2},\ref{eq3}). The weights assigned to each signal are calibrated for each country using a global optimization algorithm called differential evolution \cite{storn1997differential}.

By keeping $\beta$ and $\mu$ common across countries, we calibrate the weights ($a_i,w_i,c_i$) for every country i, considered in the analysis. We choose the combination of $\beta$ and $\mu$ that minimizes model error across countries using a global optimization algorithm (see Data and Methods for details).This gives \(\beta \approx 0.3\) and \(\mu =1\). This indicates that people reconsidered their decisions regarding masks on a daily basis. We assume $\beta$ and $\mu$ to be fixed across countries to focus on risk, policy, and norms, but they could also potentially vary across countries.

Figure \ref{fig2} shows the high variability in reported mask usage and the landscape of evolving policy mandates and disease incidence across countries. We plot these along with the calibrated model output for four countries placed in very different geographical and cultural contexts, namely the Czech Republic, Nigeria, Sweden, and South Korea. This illustrates the model’s flexibility in capturing variation in collective mask-wearing trajectories in countries facing very different disease and policy contexts. Additionally, the uncertainty in the model’s outputs, as gauged by the spread in the model output due to parameter uncertainty, was found to be negligible.

\begin{figure}[h]
\centering
\includegraphics[width=0.8\textwidth]{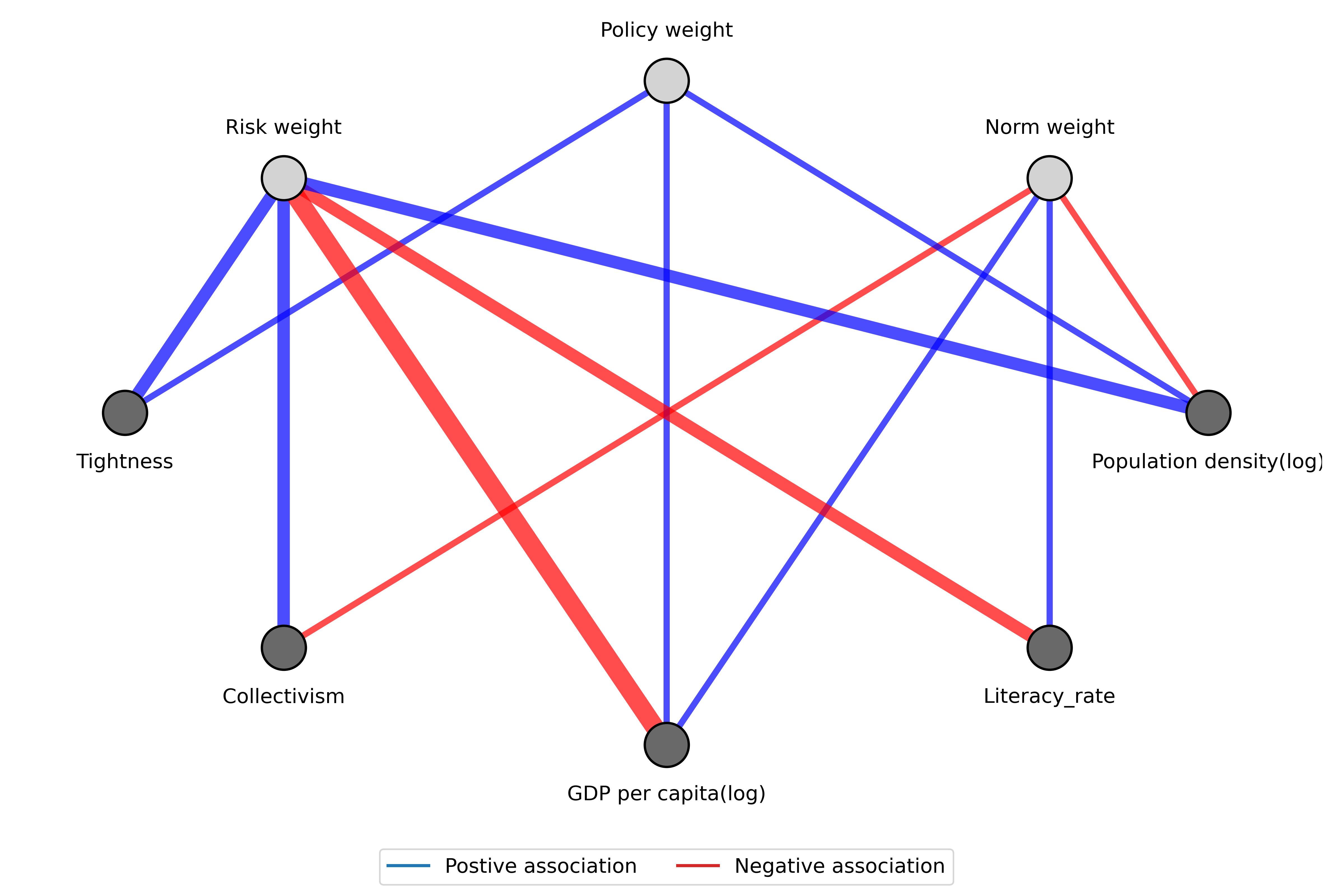}
\caption{\textbf{Significant associations of parameters with cultural and economic indicators.}
Bipartite network illustrating the relationships between socio-cultural and economic (SCE) variables (bottom row) and model parameters (top row) across five groups of countries clustered hierarchically based on SCE variables (Fig. \ref{figs7}). A link between a parameter and a variable is drawn when both show significant differences ($p\leq0.05$) (Welch’s $t$-test) in at least one cluster compared to the rest of the countries (Fig. \ref{figs8}, \ref{figs9}). This process is done for all sets of weights calibrated across bootstraps, and the aggregated links across bootstraps and clusters are then used to represent the strength of the association, depicted in the thickness of the link. Edge colors represent positive(blue) and negative association(red).} \label{fig3}
\end{figure}

The strength of the three drivers of mask usage---policy, risk, and norms---is reflected in the country-specific calibrated weights. Notably, we find that residents in African countries place a high weight on risk compared to other countries (Figure \ref{figs5} and the inset plot in Figure \ref{fig2} for Nigeria). The normalized values of the three parameters across countries are plotted in the appendix (Figures \ref{figs4},\ref{figs5},\ref{figs6}). Potential floor effects are seen in the values of risk weight and norm weight (Fig. \ref{figs6},\ref{figs11}), as we consider all weights to be positive. This can also indicate the absence of these mechanisms for these countries.

We further examined how the calibrated parameters relate to known socio-cultural and economic factors (SCE): collectivism, cultural tightness, GDP per capita, literacy rate, and population density by using a non-parametric approach. We use the natural logarithm of GDP per capita and population density in the analysis. The 47 countries are grouped using hierarchical clustering (using Ward's Method \cite{ward1963hierarchical}) based on the SCE factors. We obtain five clusters based on this analysis, which are shown in (Figure \ref{figs7} along with drivers of the bifurcation. We apply Welch’s $t$-tests by comparing each cluster with all remaining countries to identify SCE factors and parameters that are significantly different for a given cluster compared to the rest of the clusters. Figures \ref{figs8} and \ref{figs9} show the significant t-statistics for the SCE factors and the parameters across clusters.  An association between a parameter and an SCE factor is inferred if the values for both are significantly different in at least one cluster compared to the rest of the clusters. 


Cluster 1 shows a significantly higher risk weight, while clusters 3 and 5 show significantly lower risk weights. This indicates positive relationships between risk and SCE indicators like cultural tightness, collectivism, and population density, suggesting that individuals in closely knit societies may be more mindful of the risk they pose to themselves and others by not wearing a mask. In contrast, GDP per capita and literacy are negatively associated with the risk weight. We also find a negative association between total COVID-19-related fatalities in a country and its calibrated risk weight (Fig.\ref{figs11} and Table \ref{tb2}). This relationship may reflect a mechanism whereby a higher weight on risk leads to disease mitigation. Indeed, prior research has linked cultural tightness to lower deaths during the pandemic \cite{gelfand2021relationship}, and a higher weighting of risk could be the mechanism by which cultural tightness affects public health outcomes.

The weight placed on policy measures is significantly lower in countries of cluster 3, which have lower cultural tightness, lower GDP per capita, and lower population density. This could mean that policies may be less effective in culturally loose societies. Further, cluster 4 shows a significantly higher weight on social norms. These countries have lower collectivism and population density, with a higher GDP and literacy rate. Finally, we created a bipartite network to summarize and visualize the associations seen in Figure \ref{fig3} (the implementation is given in the code repository).  

The partial correlation network of SCE variables (Fig. \ref{parcorr}) reveals that key socioeconomic factors (GDP per capita and literacy) are closely aligned and inversely related to collectivism, as evident in clusters 1, 4, and 5 (Fig. \ref{figs8}). Importantly, it is along this joint axis that the risk weight parameter varies: countries with lower GDP and literacy but higher collectivism tend to exhibit stronger sensitivity to risk in mask-wearing behavior, while other associations are weak.

\begin{figure}[H]
\centering
\includegraphics[width=1\textwidth]{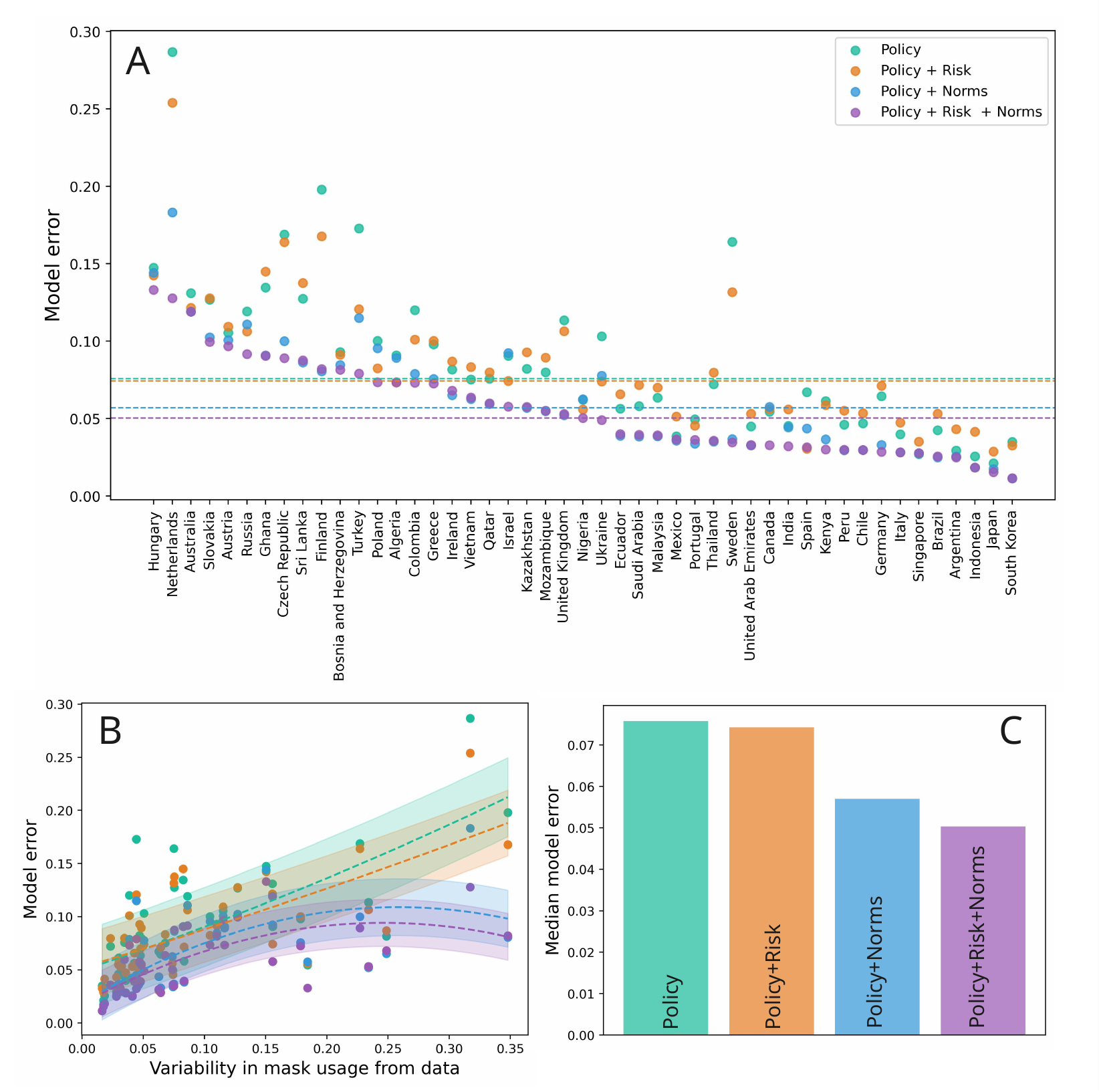}
\caption{\textbf{Building a parsimonious model:} We test the performance of different models across countries by calculating the root mean square error (RMSE) of the mask usage over time of the calibrated model against the UMD Global CTIS. In A, the model error is plotted when calibrated on 15 months of data for all countries. In B, we plot the model errors versus temporal variability in mask usage from UMD Global CTIS, which shows that the comprehensive model outperforms the other models and including conformity with social norms helps recover dynamics in countries with higher variability in mask usage. In C, the median model error across countries (shown as dotted lines in A) is plotted to highlight model performance. Table \ref{aic} shows that this insight holds when we penalize for the number of parameters.}\label{fig4}
\end{figure}

In order to determine the relative importance of the different terms for the model performance, we compared the full model to models that systematically dropped one or two of the inputs to the utility function using the Akaike Information Criterion (AIC) and the Bayesian Information Criterion (BIC), both of which penalize model complexity to avoid overfitting.  Table \ref{aic} shows that the full model consistently achieves the lowest AIC and BIC values.
This analysis underscores that the model offers a parsimonious yet effective explanation of the temporal dynamics of mask usage across 47 countries with a median error of roughly 5 percent.

The model comparisons also reveal that policy stringency is the largest determinant of mask usage: models that exclude the policy signal perform significantly worse (Figure \ref{figs1}). Focusing on the models that consider policy stringency, we find that incorporating norms to a model considering only policy leads to a significant reduction in median model error (see Fig. \ref{fig4}), and this reduction is consistent across countries (Fig.~\ref{fig4}A). Conformity is particularly important to explain mask-wearing dynamics in countries with pronounced variability in mask usage over time (Figure \ref{fig4}B). Including risk based on deaths further improves model performance (Figure \ref{fig4}C). We also tested a model where risk is based on reported COVID-19 cases and found comparable results (Fig.\ref{figs10}).

To further assess predictive performance, we calibrated each model on nine months of data and compared its predictions against observed mask usage over the following six months. Using the median RMSE across countries as a robust measure of predictive accuracy, the full model achieved the lowest calibration error (median RMSE = 0.046) and a validation error of 0.058 (Figure \ref{figs3}). The model with only policy and norms attained a lower median validation error, though at the expense of a higher calibration error. These results indicate that, for the chosen time periods, both models provide comparable predictive performance.

\section{Discussion}\label{sec12}

In times of crisis, parsimonious models are particularly valuable due to their simplicity and often better performance with regard to sensitivity to scarce data in evolving environments \cite{puy2022models} as they can offer guidance for policymakers. They must be easy to implement and adapt, and not prohibitively expensive to run. Yet, they must balance parsimony with the inclusion of key factors shaping behavior. Effective policy design must thus account for the cultural, social, and psychological processes and factors that shape both individual and collective behavior. Dynamic models are essential for understanding the emergence of collective behavior as individuals make decisions in evolving risk and policy landscapes. 

Our process-driven framework moves beyond regression-based approaches by explicitly modeling mask-wearing dynamics as the joint outcome of endogenous social norms and evolving exogenous factors such as policy and epidemiological risk. We find that both policy and norms emerged as key determinants of behavior, consistent with earlier evidence \cite{wismans2022face}. Importantly, the calibrated parameters highlight how cultural and socio-economic differences shape the relative weight that populations assign to policy, risk, and social conformity. This framework not only clarifies the role of policy stringency, norms, and risk across countries but also links these behavioral sensitivities to broader SCE factors, providing a richer account of the drivers of cross-national variation.

 Risk sensitivity is amplified in countries with higher collectivism, cultural tightness, and population density, consistent with the idea that closely knit or densely populated societies respond more strongly to epidemiological threats \cite{gelfand2021relationship,lu2021collectivism,rocklov2020high,wei2023people}. In contrast, higher literacy rates and GDP per capita are associated with lower risk weight, suggesting that in wealthier and more educated contexts, individuals may feel buffered against health risks and rely less on direct disease cues. Policy weight decreases with lower population density, GDP per capita, and cultural tightness. This pattern reflects reduced exposure in less dense settings and the weaker institutional capacity and enforcement that often accompany lower income and looser norm enforcement. One surprising result is that we find no significant relationship between norm weight and tightness looseness, a measure of the differences in the importance of adhering to social norms across countries \cite{gelfand2021relationship,yang2022sociocultural}. However, this may be in part due to the mean-field aspect of the model, which naturally does not allow us to capture within-country variation or the influence of local social structure, which may suppress the relationship between norm weight (stronger at a local level than globally) and SCE factors. Further, the potential floor effect on norm and risk weights might have impacted the ability of our analysis to link them to SCEs.

Future research should incorporate social structures such as networks, which are likely to influence perceptions of both mask usage and disease incidence. Further, introducing heterogeneity in individual weights as well as identities would help differentiate responses within countries. This is particularly relevant in countries where the COVID-19 response was politicized and mask wearing became highly partisan \cite{baxter2022local,milosh2021unmasking,kerr2021political}. Moreover, the availability of sub-national data would allow for richer spatial and regional analyses.

In this study, we focus solely on mask-wearing behavior and do not account for the influence of other mitigation strategies, such as social distancing and mobility restrictions, which directly affect the interactions between individuals, thus determining the impact of social norms on behavior by affecting the number of interactions. This could be further researched to understand the impact of online interactions once lockdowns were introduced. Since social media is polarized, the social influence was also biased by this polarization. Further, mobility restrictions and vaccinations directly determine the risk of disease incidence and, consequently, the need for masks. Additionally, exploring the coevolution of disease severity and mask usage might reveal whether or how individuals' perceptions of the benefits of mask-wearing evolve. This would also endogenize the risk from the disease, which we have considered as an exogenous factor in the current model.  Future research might also consider how delays in information (e.g., about disease severity) might impact mask wearing and disease progression. By introducing memory, agents can learn from past experiences and trends to inform their behavioral choices. Incorporating these complexities would require additional parameters, which would likely only improve the model and its insights.

Despite its simplicity, our model offers significant advantages over standard regression models, which often reveal only correlations between predictors and outcomes. By explicitly modeling the decision process, we can infer the factors driving behavior and assess how social, cultural, and economic variables may relate to these mechanisms. For example, we find that the weight placed on risk is higher in countries with higher cultural tightness and collectivism \cite{santos2011risk}. This may provide a mechanism for the finding in prior work that countries higher in tightness or collectivism had higher mask usage \cite{lu2021collectivism}. These societies may place a greater weight on risk due to considerations of not only personal risks, but also the shared risks and negative externalities associated with the absence of mask wearing. It is important to note that the parameters calibrated in our study are in the context of mask-wearing and may not directly translate to other behaviors, but the framework remains applicable \cite{michie2011behaviour}. Product prices, subsidies, damages suffered from extreme weather events, or pollution can be considered as evolving exogenous factors driving technology and behavior adoption.

Our results suggest that in countries with low risk weights, policies should focus on improving public awareness of actual risks. Countries with higher risk weights experienced fewer fatalities, highlighting that greater risk perception can enhance the impact of policy mandates. Norms also played a crucial role, creating inertia that can slow the initiation of change but also sustaining it once a new behavioral state is reached, while amplifying momentum once shifts begin. Incorporating these insights can help policymakers better anticipate the magnitude and timing of behavioral responses to policy changes, leading to more effective interventions.

Finally, although the COVID-19 pandemic introduced unprecedented challenges globally, studying the diverse behavioral responses across countries provides valuable insights into the factors shaping collective dynamics. Such understanding is not only crucial for managing future pandemics but also for fostering socially beneficial behavioral transitions to address broader issues such as climate change and environmental degradation.

\section{Data and Methods}\label{sec11}

\subsection*{Data}
The analysis presented in this study relies on different data sources. 

First, COVID-19-related death data is obtained from the World Health Organization (WHO) and accessed through the 'Our World in Data' platform \cite{owid-covid-deaths}. This dataset provides daily counts of COVID-19-related fatalities per million, which are used to estimate risk. The data is smoothed using a moving average of a window size of 7 days. 

Second, information on mask mandates is sourced from the Oxford COVID-19 Government Response Tracker \cite{hale2021global}. We use the variable \texttt{H6M$\_$Facial Coverings}, which offers data on the policy stringency on a scale from 1 to 5 with a daily frequency around the world. 

Third, we use the self-reported mask usage data, specifically \texttt{pct$\_$mc}, the weighted percentage of respondents that have reported wearing masks all or most of the time when in public mask, taken from the University of Maryland Social Data Science Center Global COVID-19 Trends and Impact Survey (UMD Global CTIS), in partnership with Facebook \cite{fan2020university} (\url{https://covidmap.umd.edu/api.html}). This survey provides estimates of mask-wearing behavior in public by inviting a representative sample of Facebook users daily. This dataset (based on survey question C5) serves as the primary behavioral outcome in our model calibration and validation. The period considered is from April 26, 2020, to July 26, 2021, which is 457 days. 276 (calibration) +181 days (validation) is the data split used for temporal validation. The model starts with $x_t=0$ to capture pre-pandemic mask usage. This analysis is done for 47 countries, the names of which can be found in Figure \ref{fig4}. Further details on the data can be found in \cite{badillo2021global}.

Finally, the values of cultural tightness and collectivism have been taken from previous literature\cite{gelfand2021relationship,gelfand2011differences,pelham2022truly}. The data on population density, GDP per capita, and literacy rate have been accessed from the 'Our World in Data' platform. 

All the data is available and described in the repository \url{https://github.com/d-mittal/Mask_wearing_COVID}. \\

\subsection*{Behavioral update}

Individuals decide to wear a mask with a probability given by the logistic function based on $U_t$:
\[
P(U_t) = \frac{1}{1 + e^{-\beta\, U_t}},\tag{2}
\label{eq2}
\]
where \(\beta\) is the behavioral sensitivity parameter that determines the steepness of the response to changes in utility. Under the assumption of a well-mixed population, the expected mask-wearing fraction on each consecutive day, $x_{t+1}$, given the previous day usage, $x_t$ is:
\[
x_{t+1} = x_t + \mu\,(P(U_t) - x_t), \tag{3}
\label{eq3}
\]
with \(\mu\) representing the fraction of the population that re-evaluates their decision on any given day. Parameters \(\beta\) and \(\mu\) are held constant across all countries. Furthermore, the parameters \(a\), \(w\), and \(c\) are assumed to be homogeneous within each country.

\subsection*{Model Calibration and Quantifying Parameter Uncertainty}

The model was calibrated for all countries using differential evolution, which uses a stochastic bounded global optimization. Fixing  $k=1 $, \(\beta \) and \(\mu\), we calibrate the remaining parameters (\(a\), \(c\), and \(w\)) for each country by minimizing the RMSE between the model’s predicted mask-wearing fraction and the observed data from UMD Global CTIS. We consider the weights (\(a\), \(c\), and \(w\)) to be positive and calibrate them within a range [$10^{-5},20$]. We vary \(\beta\) and \(\mu\) for each optimal value of the remaining parameters to identify the values that minimize the average RMSE of the model’s predictions across countries (Figure \ref{figs2}). Our analysis indicates an optimal value of approximately \(\beta \approx 0.3\) and \(\mu =1\). 

In order to assess uncertainty in the calibrated parameters, we generate 100 bootstrapped mask usage time series for each country. This is achieved by randomly sampling (with replacement) the residuals of the calibrated model. We then recalibrate the model for each bootstrap sample using the same algorithm as above. This helps estimate the uncertainty in the values of mu and beta (see Fig. \ref{figs2}) and consequently the country-specific parameters (\(a\), \(c\), and \(w\)) (see Fig.\ref{figs4},\ref{figs5},\ref{figs6}).

The hyperparameters of the differential evolution optimization and implementation details can be found in the code made available in the GitHub repository (\url{https://github.com/d-mittal/Mask_wearing_COVID}).



\backmatter

\bmhead{Supplementary information}

See attachment.


\bmhead{Acknowledgements}
DM and VVV would like to thank CSL for the useful discussions and feedback. DM would like to acknowledge the BSPL lab for valuable feedback.


\section*{Declarations}


\begin{itemize}
\item Funding: Computational Science Lab, Informatics Institute
\item Conflict of interest/Competing interests: Nothing to declare.
\item Data availability: 
All the data, model code, and analysis are made available as a GitHub repository (\url{https://github.com/d-mittal/Mask_wearing_COVID}). The static version of the code will later be turned into a Zenodo library. 
\item Author contribution:
DM, SMC, SAL, EUW, and VVV conceptualized the study. DM wrote the paper and performed the research. All authors contributed to the analysis and editing of the paper.
\end{itemize}


\bigskip
\begin{flushleft}%


\end{flushleft}
\bibliography{sn-bibliography}

\newpage
\begin{appendices}

\section{Supplementary figures and tables}\label{secA1}

\begin{figure}[htb]
\centering
\includegraphics[width=1\textwidth]{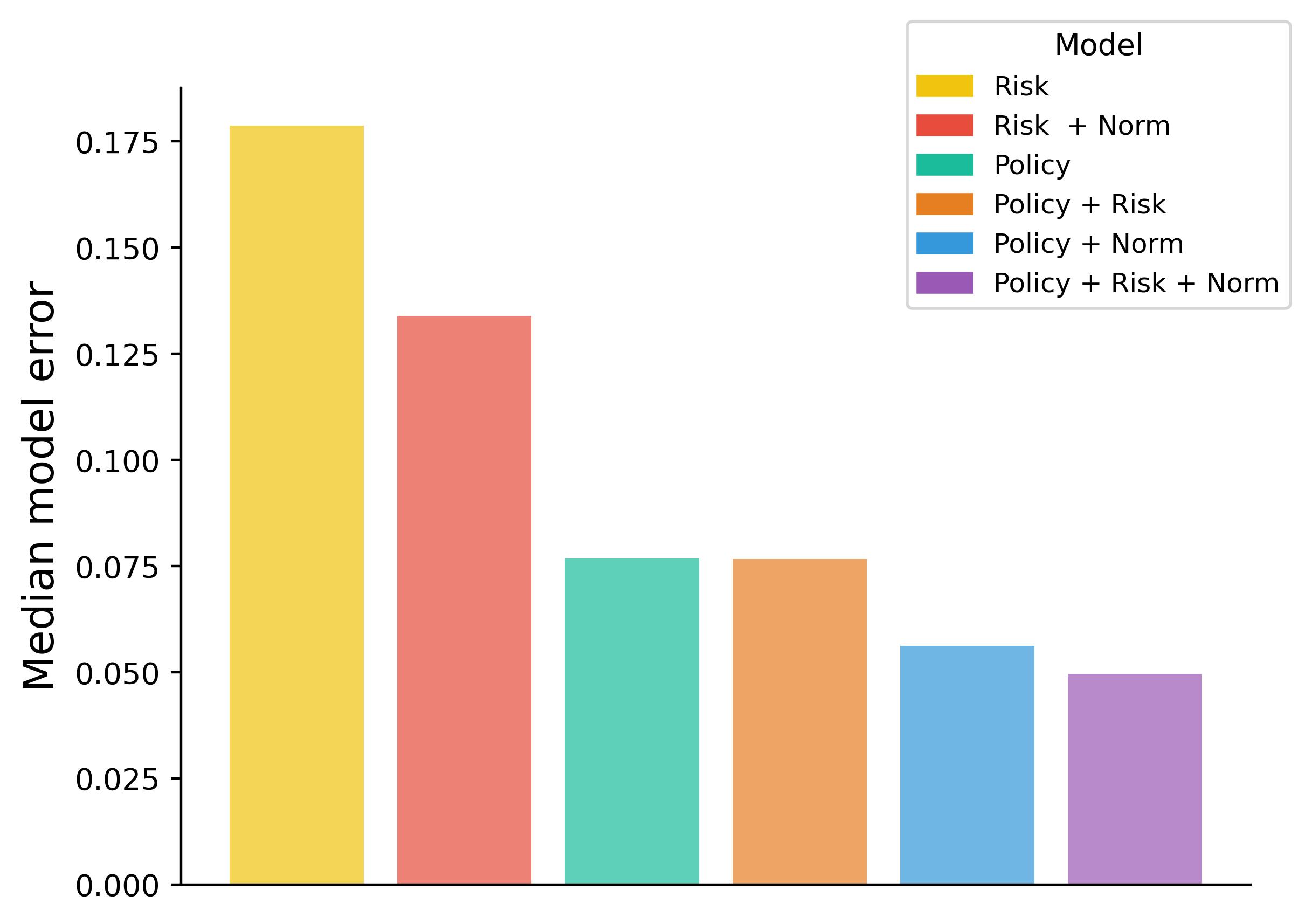}
\caption{\textbf{Model comparison.} Median model error across countries is plotted for 6 models. While policy helps explain the mask-usage greatly, conformity to norms and risk based on disease incidence help capture the data better.}\label{figs1}
\end{figure}

\begin{table}[h!]
\centering
\begin{tabular}{lcrrrr}
\toprule
\textbf{Model} & \textbf{No. of Parameters} & \textbf{AIC} & \textbf{$\Delta$AIC} & \textbf{BIC} & \textbf{$\Delta$BIC} \\
\midrule
\textbf{Policy + Risk + Norm} & \textbf{143} & \textbf{-118193.19} & \textbf{0.00} & \textbf{-117052.48} & \textbf{0.00} \\
Policy + Norm        & 96  & -112962.23 & 5230.97 & -112196.43 & 4856.05 \\
Policy + Risk        & 96  & -102763.25 & 15429.94 & -101997.46 & 15055.02 \\
Policy               & 49  & -95639.91  & 22553.28 & -95249.04  & 21803.44 \\
Risk + Norm          & 96  & -71441.97  & 46751.23 & -70676.18  & 46376.31 \\
Risk                 & 49  & -61414.03  & 56779.17 & -61023.15  & 56029.33 \\
\bottomrule
\end{tabular}
\caption{\textbf{Model comparison using AIC and BIC}, based on residual sum of squares computed over all 21,526 data points (47 countries $\times$ 458 time points). Total parameter count includes 2 global parameters and 47 country-specific parameters per the included behavioral mechanism. The comprehensive model remains decisively the best-fitting model.}
\label{aic}
\end{table}

\begin{figure}[htb]
\centering
\includegraphics[width=0.7\textwidth]{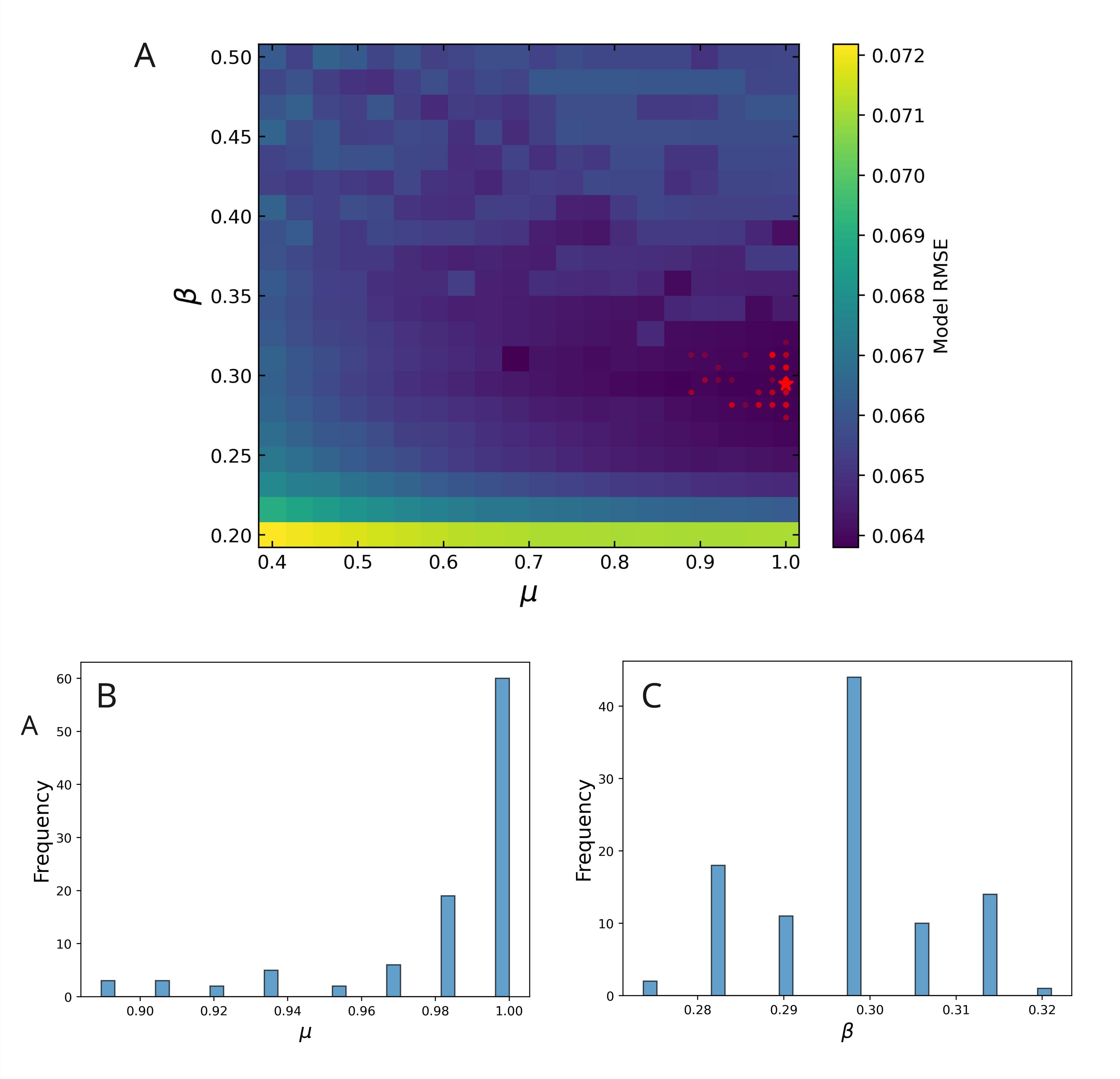}
\caption{\textbf{Fixing $\beta$ and $\mu$ across countries.} Global optimization of $a,c,$ and $w$ parameters is done for a range of values of $\beta$ and $\mu$, such the model RMSE across countries is minimized. Here we show the heatmap of model error, which gives us $\beta\approx0.3$ and $\mu=1$ depicted with a red star. The red scatter points show the values of $\mu$ and $\beta$ corresponding to the bootstraps, and panels B, C show the distribution of values of $\mu$ and $\beta$, respectively. }\label{figs2}
\end{figure}

\begin{figure}[htb]
\centering
\includegraphics[width=0.7\textwidth]{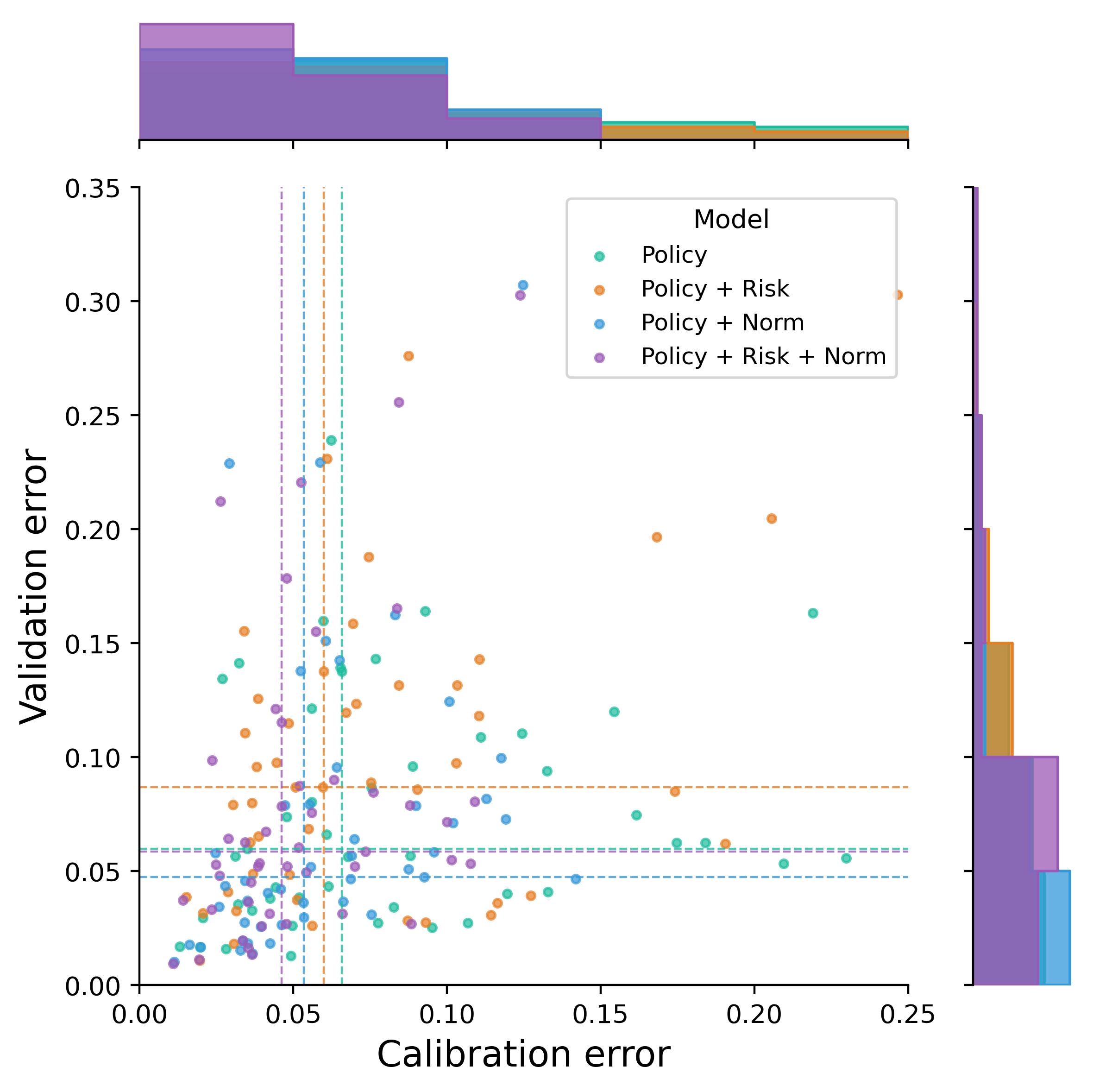}
\caption{\textbf{Temporal validation:} The models are calibrated on 9 months of data and validated against the following 6 months of data. The calibration and validation errors are plotted for 4 models. The horizontal (vertical) dashed lines depict the median validation (calibration) errors. The comprehensive model (depicted in violet) achieved the lowest calibration error. The model with only policy and norms attained a lower median validation error, though at the expense of a higher calibration error. These results indicate that, for the chosen time periods, both models provide comparable predictive performance.}\label{figs3}
\end{figure}

\begin{figure}[htb]
\centering
\includegraphics[width=1\textwidth]{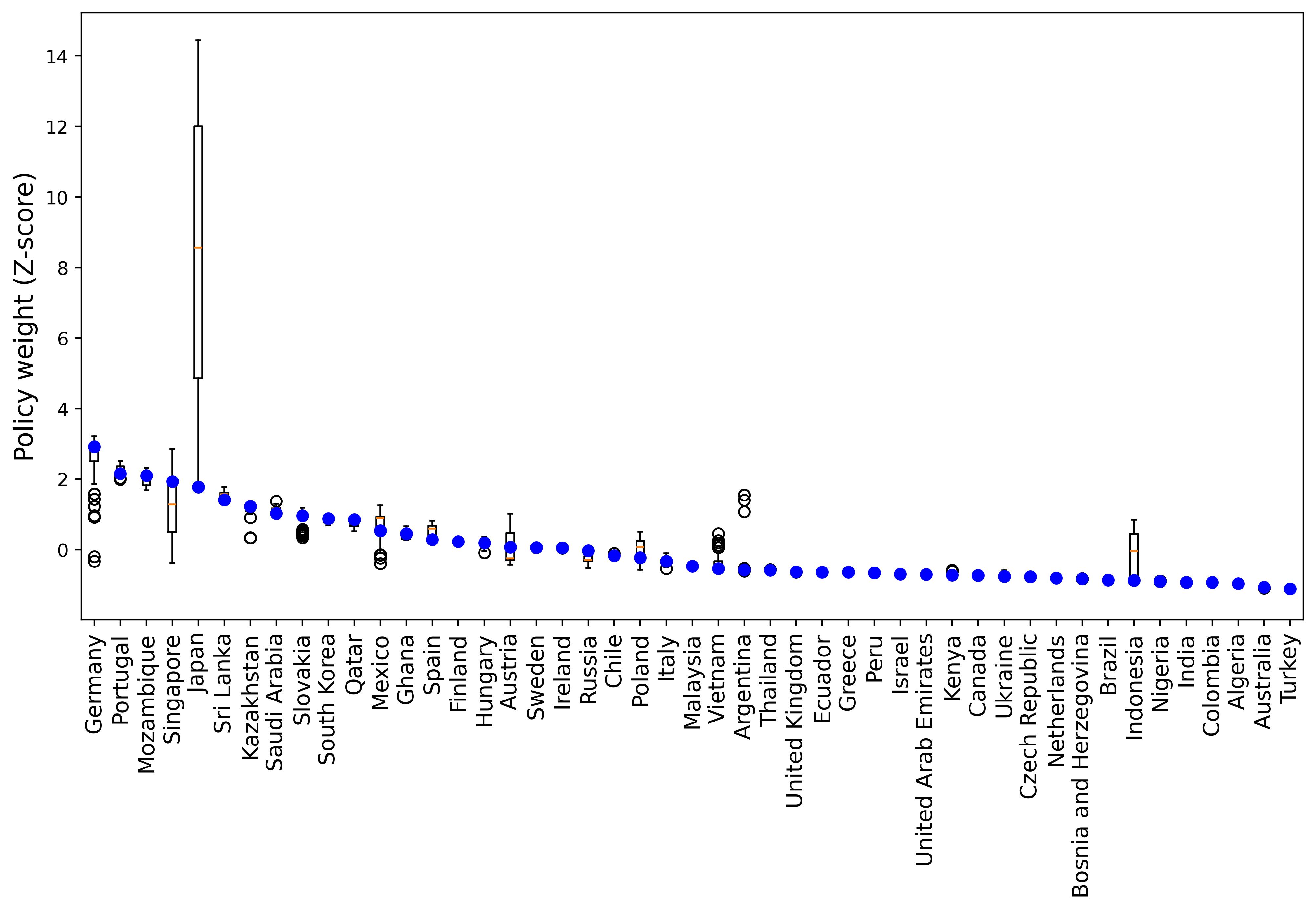}
\caption{\textbf{Policy weight:} The calibrated value of policy weight, along with the parameter uncertainty, is plotted in descending order. The box plots show the calibrated values from the bootstraps shown in Fig.\ref{figs2}.  }\label{figs4}
\end{figure}

\begin{figure}[htb]
\centering
\includegraphics[width=1\textwidth]{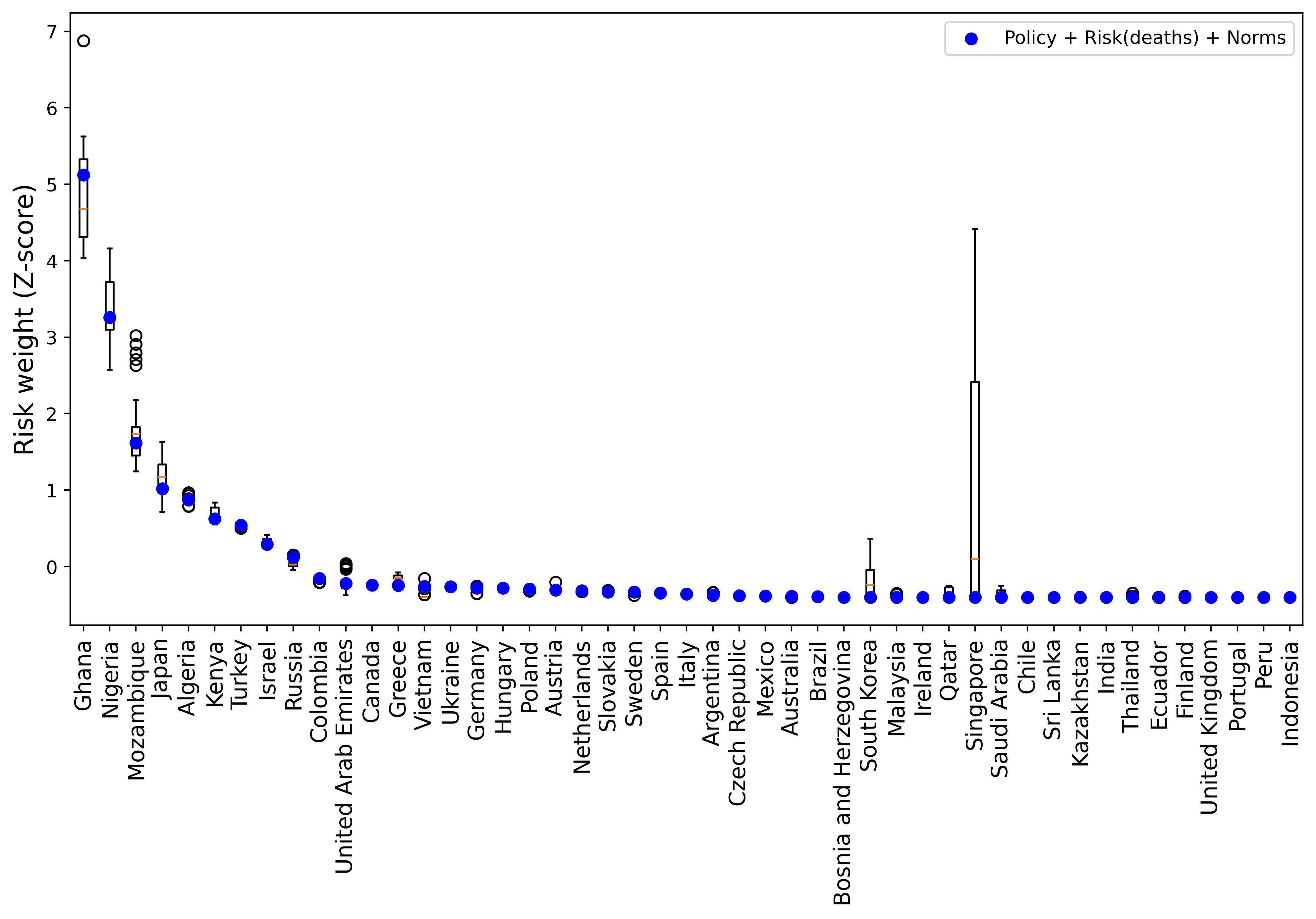}
\caption{\textbf{Risk based on disease incidence:} The calibrated value of risk weight, along with the parameter uncertainty, is plotted in descending order. We see that African countries show the highest risk weight based on disease incidence, while it remains negligible for most of the countries. There may be flooring effects for some of the countries in cluster 2 (see Fig.~\ref{figs11}). The box plots show the calibrated values from the bootstraps shown in Fig.\ref{figs2}. }\label{figs5}
\end{figure}

\begin{figure}[htb]
\centering
\includegraphics[width=1\textwidth]{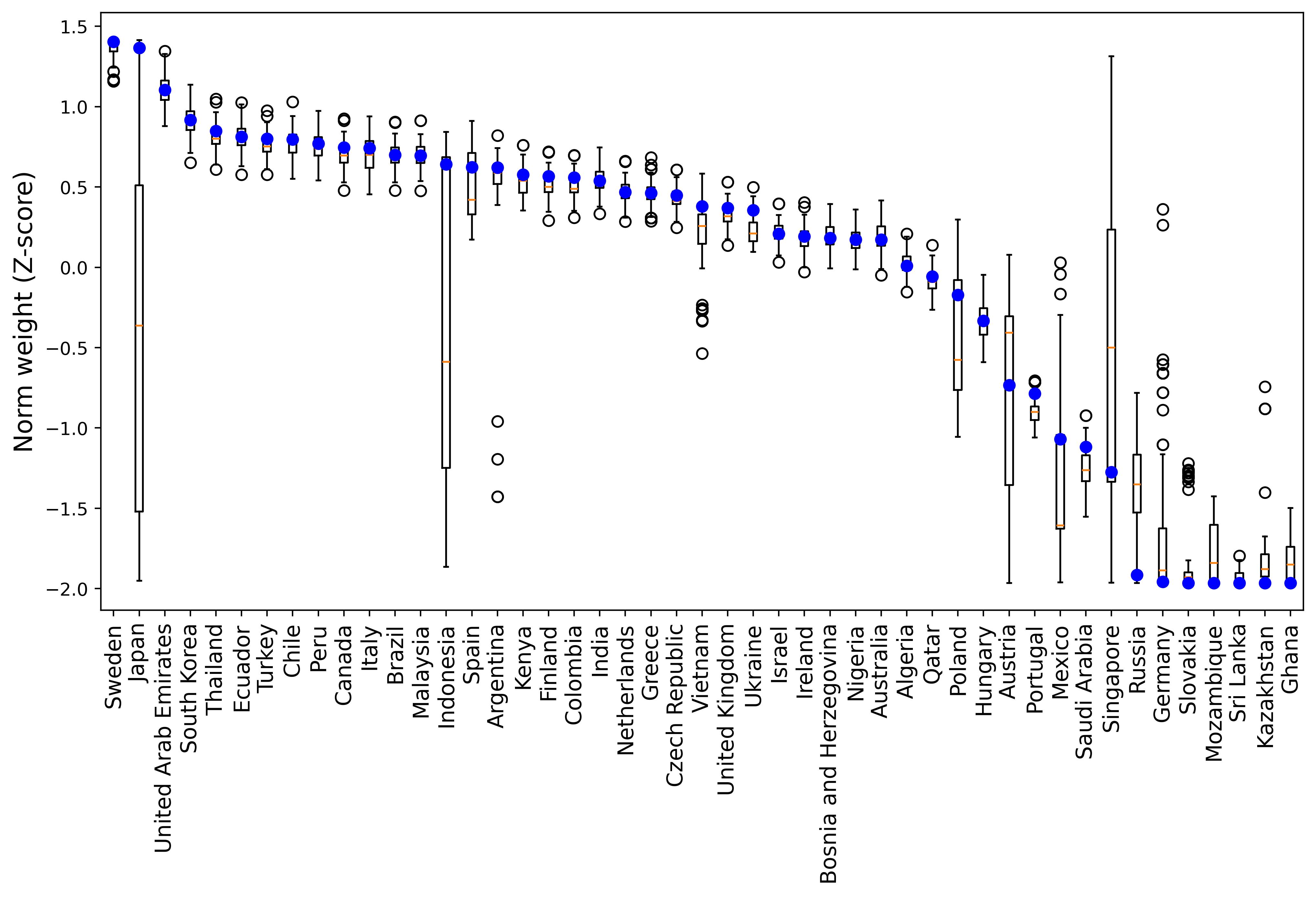}
\caption{\textbf{Norm weight:} The calibrated value of norm weight, along with the parameter uncertainty, is plotted in descending order. The parameter remains comparable across countries, barring a few exceptions, which indicate a flooring effect. The box plots show the calibrated values from the bootstraps shown in Fig.\ref{figs2}. }\label{figs6}
\end{figure}

\begin{figure}[htb]
\centering
\includegraphics[width=1\textwidth]{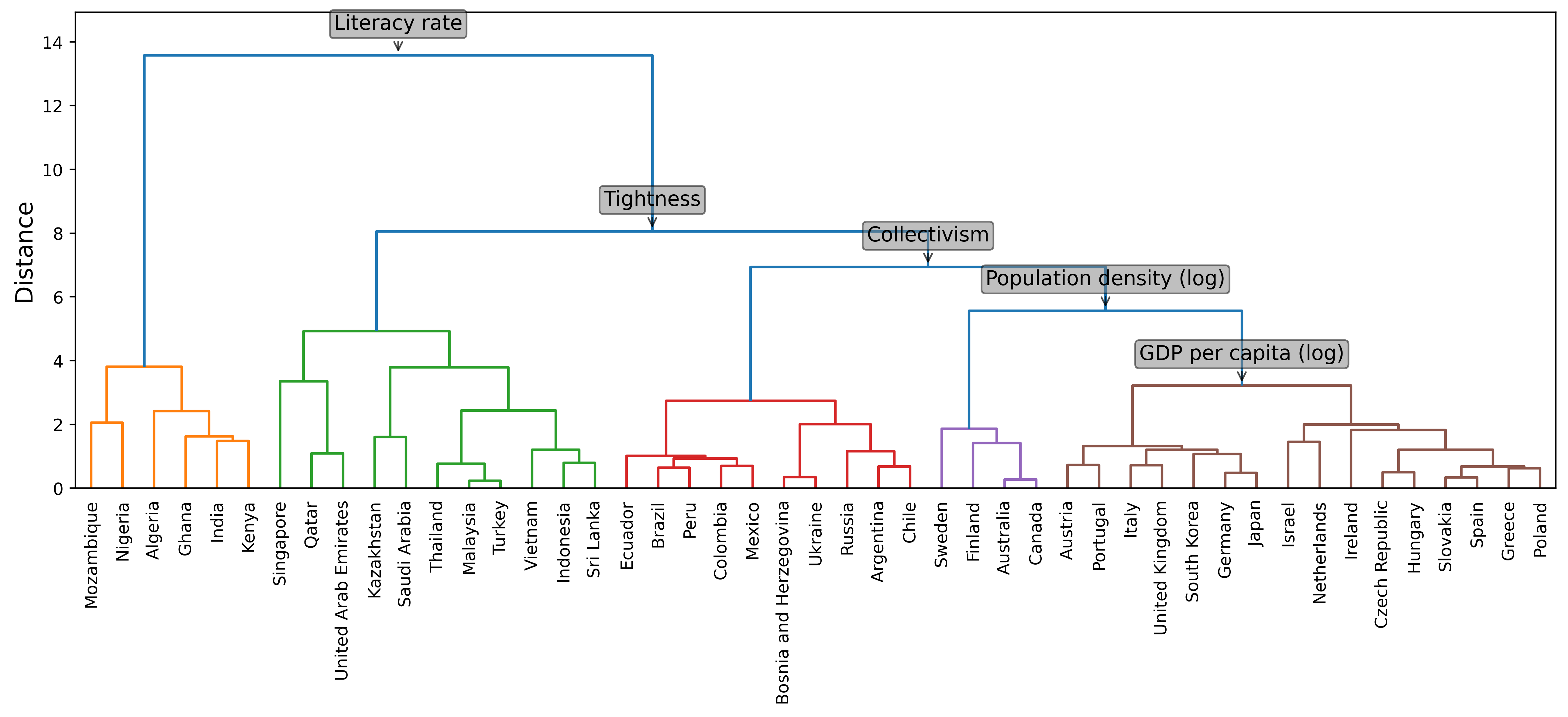}
\caption{\textbf{Hierarchical clustering of countries} based on socio-cultural and economic factors: Cultural tightness, collectivism, GDP per capita, literacy rate, and population density. The threshold distance is set such that 5 distinct clusters emerge. }\label{figs7}
\end{figure}

\begin{figure}[htb]
\centering
\includegraphics[width=0.7\textwidth]{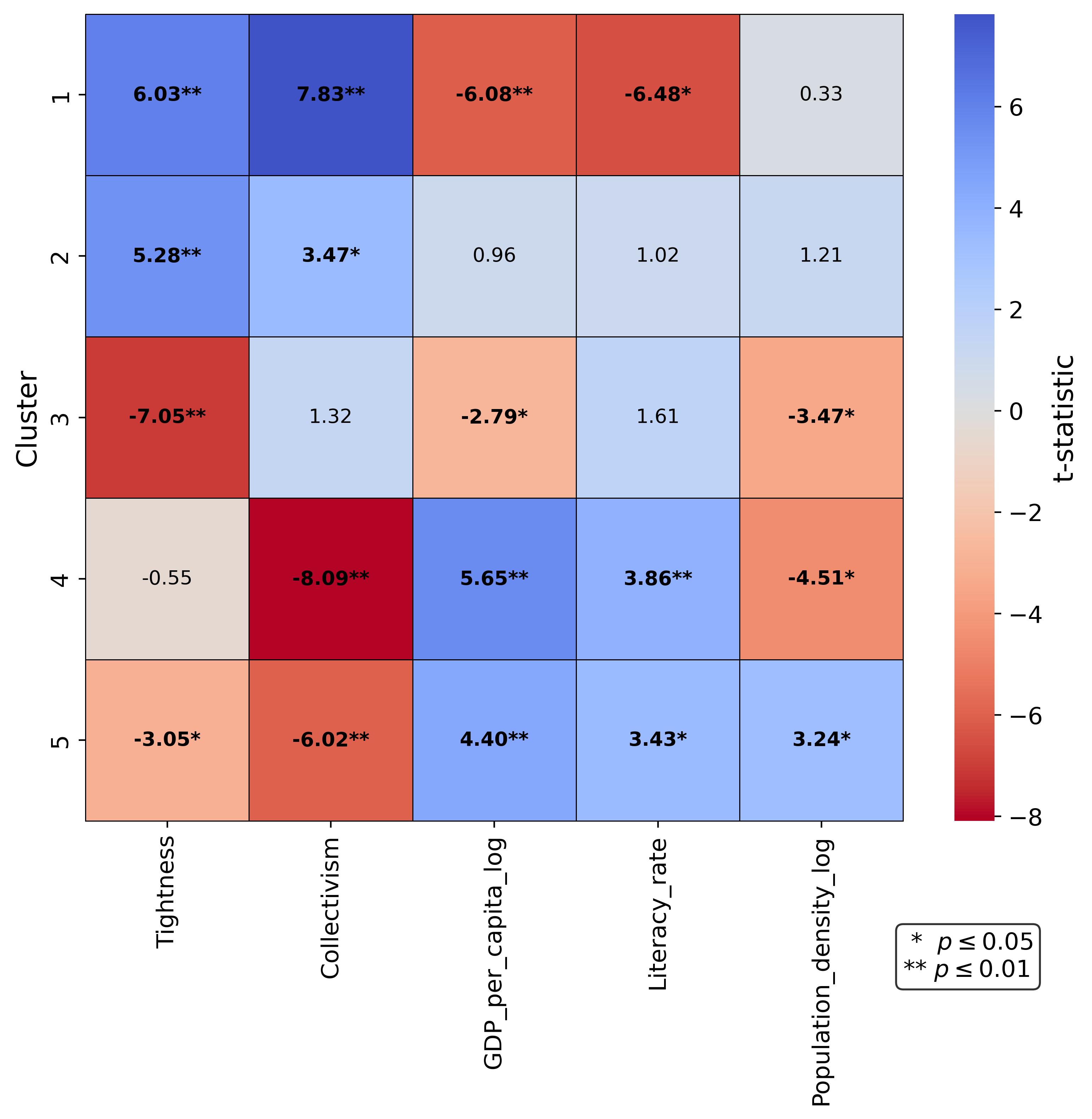}
\caption{\textbf{Characteristic SCE factors within clusters } We perform two-tailed Welch’s $t$-tests to identify SCE factors that differ significantly between each cluster and the remaining countries. The color scale indicates whether the factor values are significantly higher (blue) or lower (red) for the cluster compared to the rest of the population. Stars indicate the level of significance.}\label{figs8}
\end{figure}

\begin{figure}[h]
\centering
\includegraphics[width=0.7\textwidth]{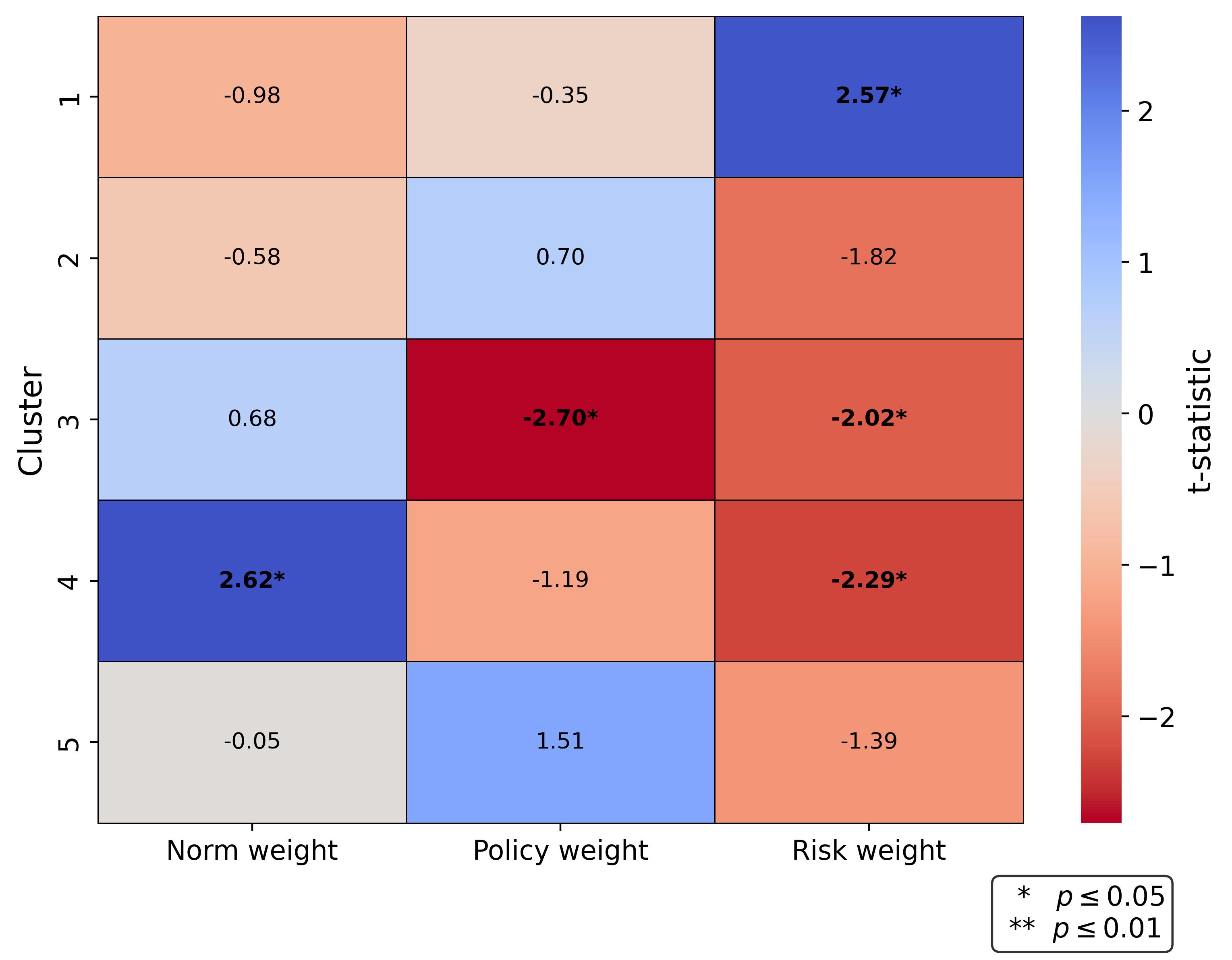}
\caption{\textbf{Characteristic parameters within clusters }We perform two-tailed Welch’s $t$-tests to identify calibrated parameters that differ significantly between each cluster and the remaining countries. The color scale indicates whether the factor values are significantly higher (blue) or lower (red) for the cluster compared to the rest of the population. Stars indicate the level of significance.} \label{figs9}
\end{figure}

\begin{figure}[h]
\centering
\includegraphics[width=0.6\textwidth]{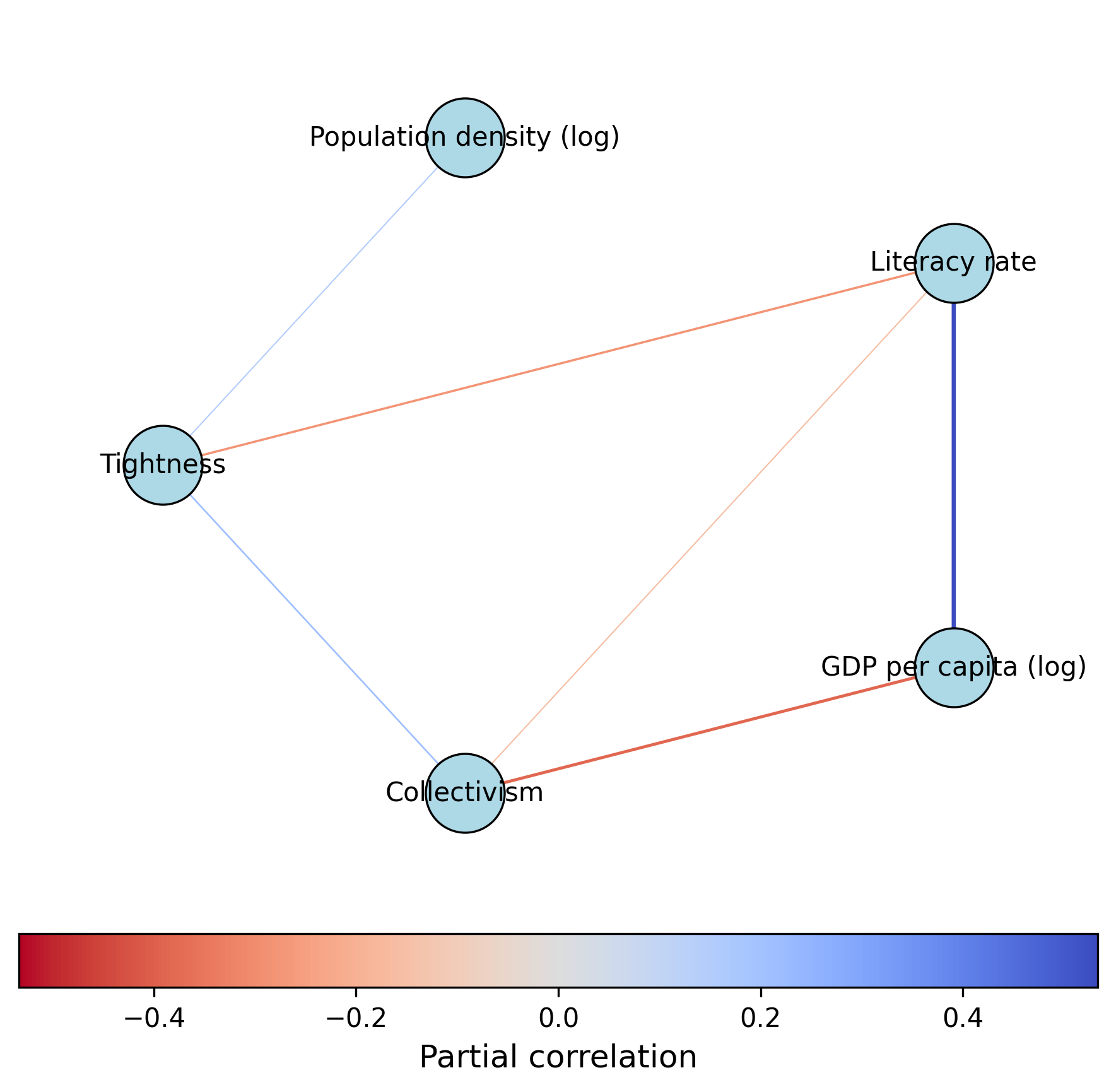}
\caption{\textbf{Partial correlation network of socio-cultural and economic (SCE) indicators} estimated with Graphical Lasso and cross-validated regularization on country-level data. Edges represent conditional associations between indicators after controlling for all others. Only a few robust links were retained, with the strongest being a positive relation between GDP per capita and literacy rate and a negative relation between GDP per capita and collectivism. Edge color indicates the sign and thickness reflects the magnitude of the partial correlation.} \label{parcorr}
\end{figure}

\begin{figure}[h]
\centering
\includegraphics[width=1\textwidth]{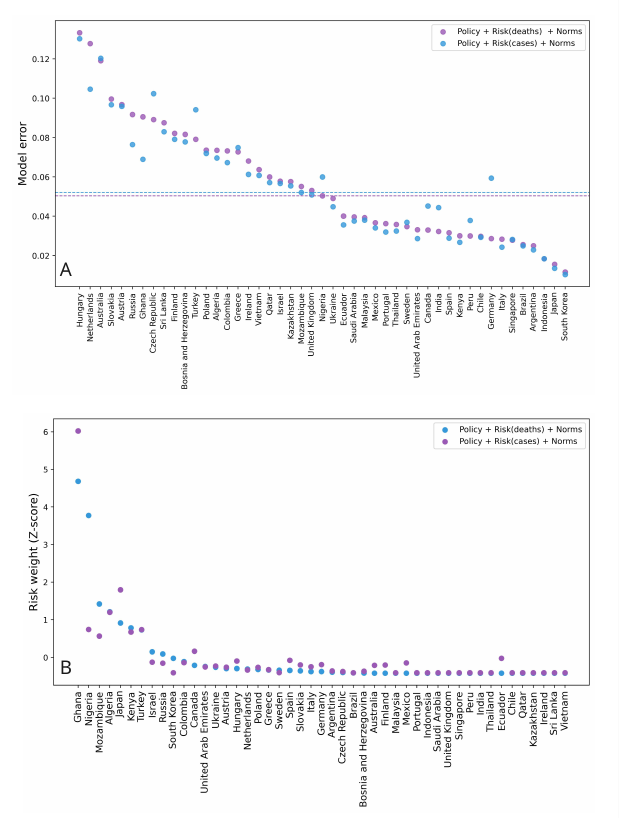}
\caption{\textbf{Comparison of models with risk based on COVID-19 cases and deaths.} A)Model errors are plotted for models with either basis of risk. We find minimal differences, which is expected given the high correlation between the number of cases and the number of deaths. The dotted line shows the median model error. B) The z-scored values of risk weights for both models are plotted, showing a high Pearson correlation of 0.86 ($p<<0.001$). }\label{figs10}
\end{figure}

 \FloatBarrier

\begin{figure}[tp]
\centering
\includegraphics[width=0.6\textwidth]{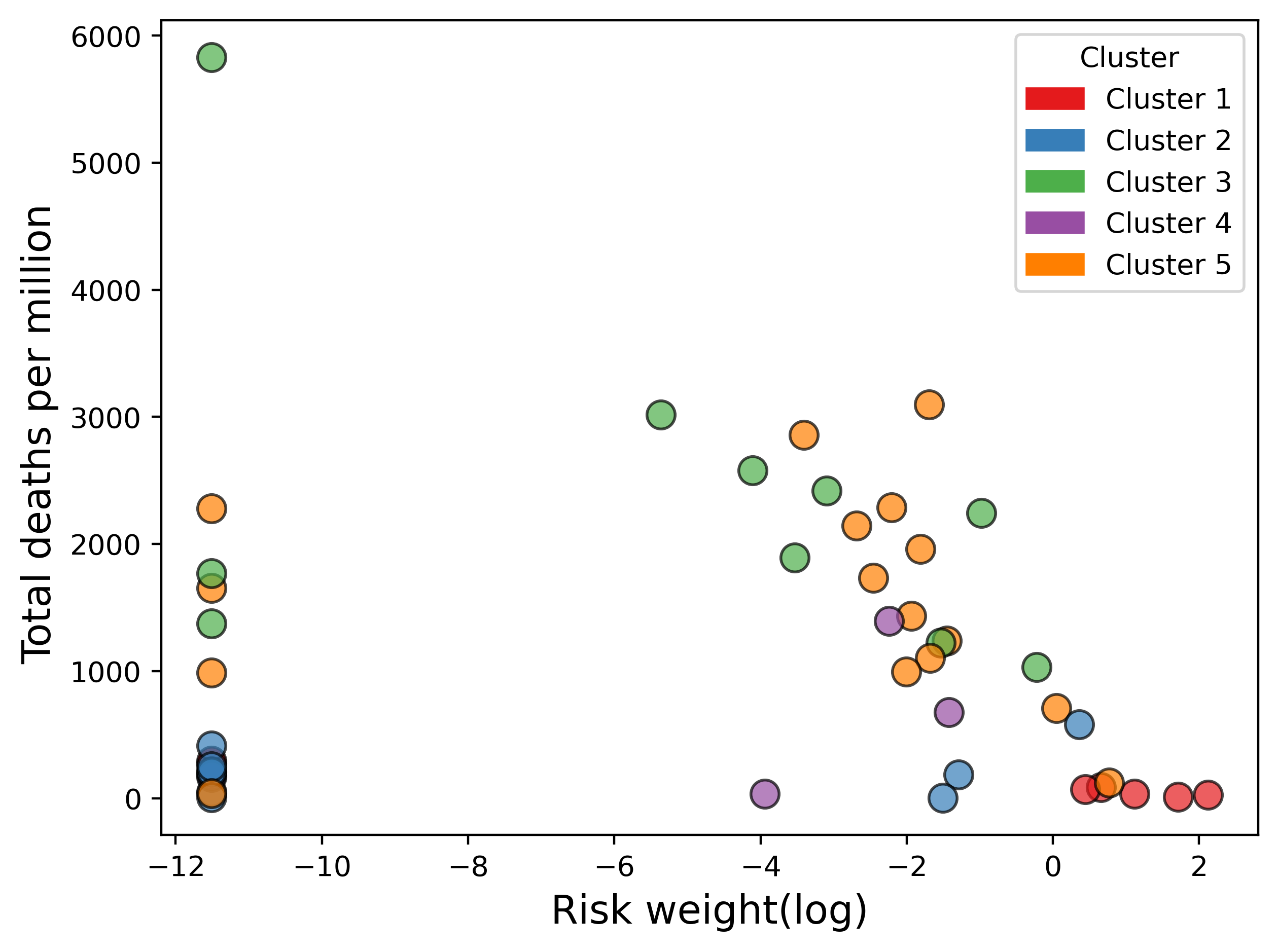}
\caption{\textbf{Total number of deaths per million vs. calibrated risk weight.} This plot shows that countries with a greater number of COVID-19-related fatalities showed lower risk weights. For this plot, the log of the risk weight is used. The colors show the clusters obtained in Fig. \ref{figs7}. }\label{figs11}
\end{figure}

\FloatBarrier

\begin{table}[h]
\centering
\caption{OLS regression results: association between weights and total deaths per million (standardized values)}
\begin{tabular}{lcccc}
\hline
\textbf{Variable} & \textbf{Coefficient} & \textbf{Std. Error} & \textbf{t} & \textbf{p-value} \\
\hline

Intercept      & $2.81 \times 10^{-16}$ & 0.141  & $1.99 \times 10^{-15}$ & 1.000 \\
\textbf{Risk weight}    & \textbf{-0.318} & \textbf{0.149} & \textbf{-2.138} & \textbf{0.038} \\
Policy weight  & -0.238 & 0.187 & -1.277 & 0.208 \\
Norm weight    & -0.105 & 0.192 & -0.547 & 0.587 \\
\hline
\multicolumn{5}{l}{$R^2$ = 0.127, Adjusted $R^2$ = 0.066, No. Observations = 47} \\
\multicolumn{5}{l}{F-statistic = 2.087, Prob(F-statistic) = 0.116, AIC = 134.0, BIC = 141.4} \\
\hline
\end{tabular}\label{tb2}
\end{table}

\end{appendices}



\end{document}